\documentclass[aps,twocolumn, floatfix,superscriptaddress]{revtex4}
\usepackage{graphicx}
\usepackage{hyperref}
\usepackage{amsmath}
\usepackage{amsfonts}
\usepackage{mathtools,bm}
\usepackage[amssymb]{SIunits}
\usepackage{color}
\usepackage{mathrsfs}

\newcommand{\ket}[1]{|{#1}\rangle} 
\newcommand{\bra}[1]{\langle #1 | }

\newcommand{\kket}[1]{\left|\hskip-.02cm\left| #1\right\rangle\hskip-.05cm\right\rangle}

\DeclareMathOperator{\e}{e}

\begin{document}
\flushbottom
\title{Self-interfering wavepackets}

\author{David Colas}
\affiliation{Departamento de F\'isica Te\'orica de la Materia Condensada and Condensed Matter Physics Center (IFIMAC), Universidad Aut\'onoma de Madrid, E-28049, Spain}

\author{Fabrice P. Laussy}
\affiliation{Departamento de F\'isica Te\'orica de la Materia Condensada and Condensed Matter Physics Center (IFIMAC), Universidad Aut\'onoma de Madrid, E-28049, Spain}
\affiliation{Russian Quantum Center, Novaya 100, 143025 Skolkovo, Moscow Region, Russia}

\begin{abstract}  
  We study the propagation of non-interacting polariton wavepackets.
  We show how two qualitatively different concepts of mass that arise
  from the peculiar polariton dispersion lead to a new type of
  particle-like object from non-interacting fields---much like
  self-accelerating beams---shaped by the Rabi coupling out of
  Gaussian initial states.  A divergence and change of sign of the
  diffusive mass results in a ``mass wall'' on which polariton
  wavepackets bounce back. Together with the Rabi dynamics, this yield
  propagation of ultrafast subpackets and ordering of a spacetime
  crystal.
\end{abstract}

 \date{\today} \maketitle

Field theory unifies the concepts of waves and
particles~\cite{ng_book09a}. In quantum physics, this brought at rest
the dispute of the pre-second-quantization era, on the nature of the
wavefunction. As one highlight of this conundrum, the coherent state
emerged as an attempt by Schr\"odinger to prove Heisenberg that his
equation is suitable to describe particles since some solutions exist
that remain localized~\cite{schrodinger26c}
. However, the reliance on an external potential and the lack of other
particle properties---like resilience to collisions---makes this
qualification a moot point and quantum particles are now understood as
excitations of the field.  The deep connection between fields and
particles is not exclusively quantum and classical fields also provide
a robust notion of particles, most famously with
solitons~\cite{korteweg95a}. The particle cohesion is here assured
self-consistently by the interactions, allowing free propagation and
surviving collisions with other solitons (possibly with a phase
shift).  For a long time, this has been the major example of how to
define a particle out of a classical field, until Berry and Balazs
discovered the first case of a similar behaviour in a non-interacting
context: the Airy beams~\cite{berry79a}. These solutions to
Schr\"odinger equation (or equivalently through the Eikonal
approximation, to Maxwell equations) retain their shape as they
propagate as a train of peaks (or sub-packets) and also exhibit
self-healing after passing through an obstacle~\cite{broky08a}. The
ingredient powering these particle behaviours is phase-shaping,
assuring the cohesion by the acceleration of the sub-packets inside
the mother packet.  The solution was first regarded as a mathematical
curiosity as it is not normalizable,
till a truncated version was experimentally realized and shown to
exhibit this dramatic phenomenology but for a finite
time~\cite{siviloglou07a}. The Airy beam is now a recognized
particle-like object, in some cases emerging from fields that quantize
elementary particles~\cite{voloch13a}, thus behaving like a
meta-particle.  It is in fact but one example of a full family of
so-called ``accelerating beams''~\cite{zhang12a}, that all similarly
endow linear fields with particle properties: shape-preservation and
resilience to collisions.

\begin{figure}[t]
  \includegraphics[width=\linewidth]{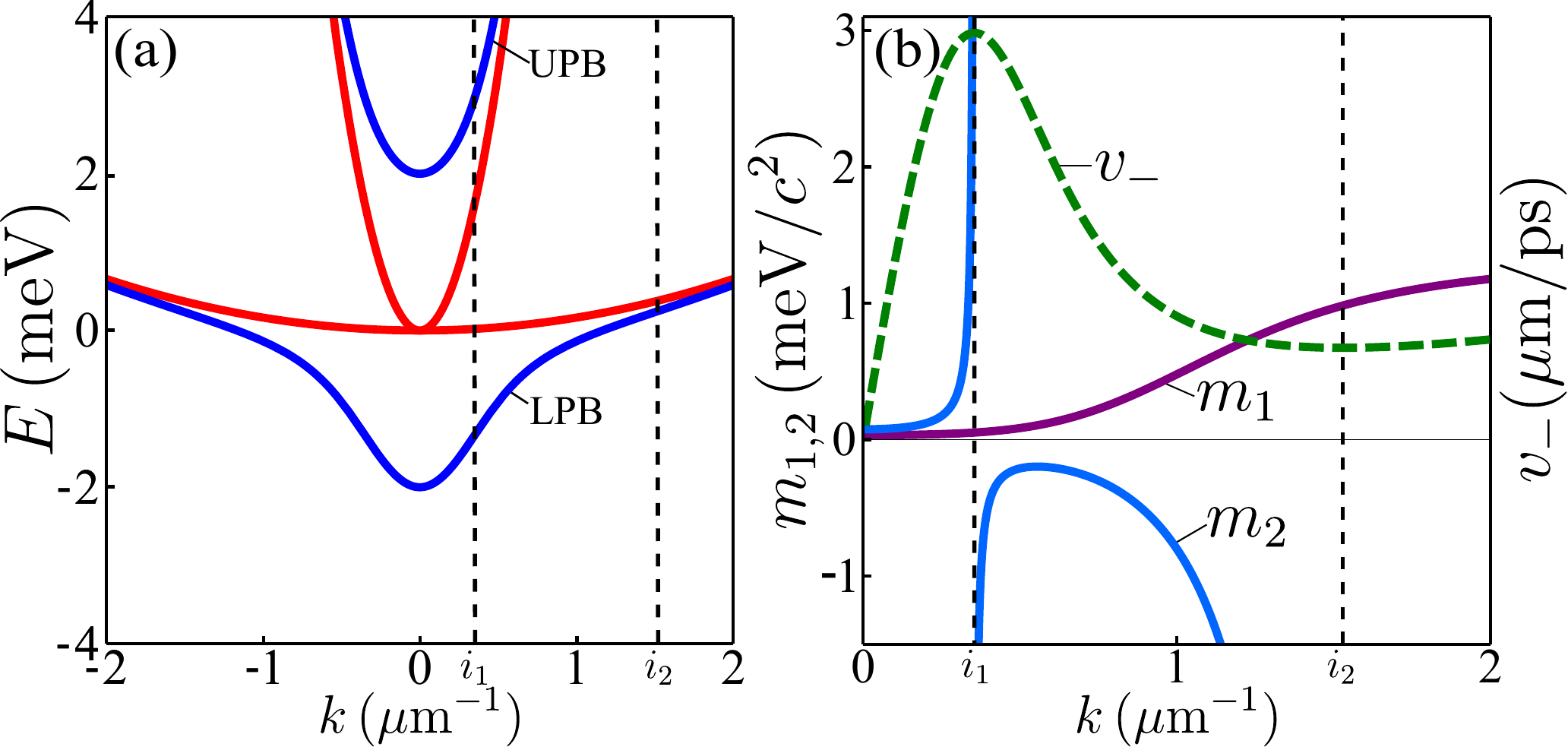}
  \caption{(Color online) a) Polariton dispersions. In red: the
    parabolic dispersion of the cavity photon, and the bare
    exciton. In blue: the polariton branches~$E_\pm$. The vertical
    dashed lines at $i_1$ and $i_2$ mark the inflexion points of the
    LPB.  b) Effective masses for the LPB as a function of momentum:
    in purple the inertial mass $m_1$, in blue the diffusive mass
    $m_2$ (negative when~$i_1<k<i_2$), in green the group velocity
    $v_-$.}
  \label{fig:1}
\end{figure}

\begin{figure}[t]
  \includegraphics[width=1\linewidth]{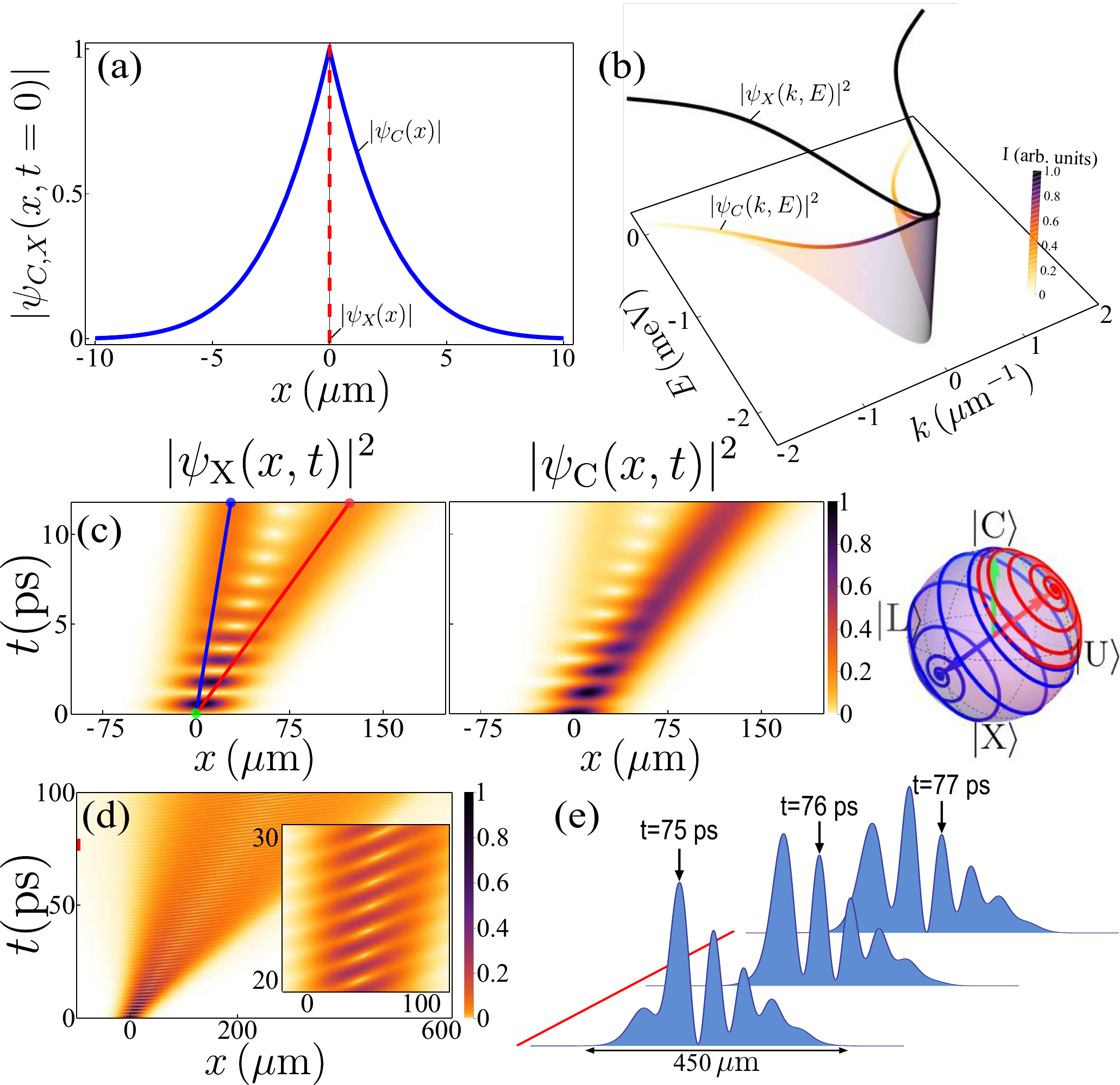}
  \caption{(a) Localizing a polariton in space is possible for one of
    its component only (here the one in dashed red); the other field
    smears out to keep the particle on its branch. (b) Counterpart
    of~(a) in energy-momentum space with forced localization of the
    photon field by the delocalized exciton.  (c) Spacetime evolution
    of $|\psi_{\textrm{C}}(x,t)|^2$ and $|\psi_{\textrm{X}}(x,t)|^2$
    with as an initial condition a photon of momentum
    $k_0=\unit{0.5}{\per\micro\meter}$. The Bloch Sphere shows the
    quantum state trajectories along the line of the density plot
    in~(c).  (d) Configuration with $\Delta=-\Omega_\mathrm{R}$,
    preventing splitting of the beam and resulting in ultrafast,
    Rabi--powered, propagating sub-packets, as shown for three
    snapshots of time in~(e).}
  \label{fig:2}
\end{figure}

In this Letter, we add another member to the family of mechanisms that
provide non-interacting fields with particle properties. Namely, we
show that two coupled fields of different masses can support
self-interfering wavepackets, resulting in the propagation of a train
of sub-packets, much like the Airy beam, but without acceleration,
fully-normalizable and self-created out of a Gaussian initial
state. Such coupled fields can be conveniently provided in the
laboratory by polaritons~\cite{kavokin_book11a}, the quantum
superposition of the spatially extended light~$\psi_\mathrm{C}(x,t)$
and matter fields~$\psi_\mathrm{X}(x,t)$, cf.~Fig~\ref{fig:1}.  They
find their most versatile and tunable implementation in semiconductors
where excitons (electron-hole pairs) of a quantum well are coupled to
the photons of a single-mode of a microcavity.  We will consider the
simplest~1D case, realized for instance in quantum
wires~\cite{wertz10a} but similar results hold for the more common
planar geometry. Since polaritons can form condensates giving rise to
a wavefunction that describes their collective
dynamics~\cite{kasprzak06a}, they are a dream laboratory to
investigate the wavepacket propagation in a variety of
contexts~\cite{sanvitto10a}, such as propagation of
spin~\cite{shelykh06a}, bullets~\cite{amo09a} or Rabi
oscillations~\cite{liew14a} with technological applications already in
sight~\cite{ballarini13a,espinosa13a}. 
Polaritons are highly valued for their nonlinear properties due to the
particles self-interactions~\cite{carusotto13a}, illustrated by a
whole family of solitons (bright, dark,
composite\dots)~\cite{egorov09a,sich12a,flayac11a,christmann14a}. Recently,
however, also the non-interacting regime has proved to be topical,
with reports of skyrmions analogues\cite{vishnevsky13a}, band
structure engineering~\cite{jacqmin14a} focusing and conical polariton
diffraction~\cite{tercas14a}, internal Bosonic Josephson
junctions~\cite{arXiv_voronova15a}, emulates of oblique dark and half
solitons~\cite{cilibrizzi14a
}, $\mathbb{Z}$ topological insulator~\cite{nalitov15a} or the
implementation of Hebbian learning in neural
networks~\cite{espinosa15a} to name a few but illustrative
examples. In most of these cases, interactions bring the physics to
even farther extents rather than spoiling the underlying linear
effect, that remains nevertheless the one capturing the
phenomenon. The linear regime can be achieved at low
densities~\cite{dominici14a} since the polariton interaction at the
few particles level is small. In this case, the dynamics of the
wavefunction~$\ket{\psi}$ is ruled by the polariton propagator~$\Pi$
such that~$\ket{\psi(t)}=\Pi(t-t_0)\ket{\psi(t_0)}$. In free space,
the propagator is diagonal in~$k$ space~\cite{carusotto13a}:
\begin{equation}
  \label{eq:lunjun29102131CEST2015}
  \bra{k'}\Pi(t)\ket{k}=\exp\left[-i
    \begin{pmatrix}
      \frac{\hbar k^2}{2 m_\mathrm{C}}+\Delta & \Omega_\mathrm{R} \\
      \Omega_\mathrm{R} & \frac{\hbar k^2}{2m_\mathrm{X}} \\
    \end{pmatrix}
    t\right]\delta(k-k')
\end{equation}
where $m_\mathrm{C}$ is the photon mass, $m_\mathrm{X}$ the exciton
mass, $\Delta$ their detuning and $\Omega_\mathrm{R}$ their Rabi
coupling. The eigenstates of the propagator,
$\Pi(t)\kket{k}_\pm=\exp(-iE_\pm t)\kket{k}_\pm$, define both the
polariton dispersion~$E_\pm=\hbar k^2 m_+ +2\Delta \mp k_\Omega^2$ and
the canonical polariton basis
$\kket{k}_\pm\propto(E_\pm(k),1)^T\ket{k}$ where~$m_\pm=(m_\mathrm{C}
\pm m_\mathrm{X})/(m_\mathrm{C}m_\mathrm{X})$ are the reduced relative
masses, $k_\Omega=\sqrt[4]{\hbar^2 k^4 m_-^2 -4\hbar k^2 \Delta m_-
  +4(\Delta^2+4\Omega_\mathrm{R}^2)}$ the dressed momentum and
$\ket{k}$ the plane wave of well-defined momentum~$k$. We use the
notation~$\kket{}_\pm$ for upper~($+$) and lower~($-$) polaritons. A
general polariton state is thus expressed as
$\kket{\psi}_\pm=\int_{-\infty}^{\infty}\phi_\pm(k)\kket{k}_\pm\,dk$
where $\phi_\pm(k)$ is the scalar-field polariton wavefunction.
Except for a well-defined polariton state in~$k$-space, i.e., a fully
delocalized polariton in real space, the photon and exciton components
of a polariton cannot be jointly defined according to a given
wavepacket~$\phi(k)$.  Indeed, except if~$\phi(k)=\delta(k)$, one
component gets modulated by the $E_\pm(k)$ factor needed to maintain
the particle on its own branch. One striking consequence of this
composite structure is that a polariton cannot be localized in
real-space, in the sense that both its photon and exciton components
be simultaneously localized. Choosing~$\phi(k)$ such that
either~$\psi_\mathrm{C}(x,t=0)$ or $\psi_\mathrm{X}(x,t=0)$
is~$\delta(x)$ results in smearing out the other component in a
pointed wavefunction surrounding the singularity of the localized
field, as shown in Fig.~\ref{fig:2}(a--b). Such constrains result in a
rich phenomenology when involving a large enough set of momenta which,
to the best of our knowledge, remained up to now safely hidden behind
the simplicity of the problem.  We devote the rest of the text to some
of these remarkable effects, arising from the self-shaping and
self-interferences of polaritons due to their composite structure,
always in a non-interacting context.

It has long been known that the mass imbalance~$m_\mathrm{C}\ll
m_\mathrm{X}$ gives rise to a peculiar dispersion relation for the
upper ($E_+$) and lower ($E_-$) polariton branches, shown in
Fig.~\ref{fig:1}(a) in blue along with, in red, the parabolic
dispersions of the light photon and the heavy exciton, meeting
at~$k=0$ ($\Delta=0$). The dynamics of a Gaussian wavepacket that is
large enough in space to probe only parabolic portions of the
dispersion in reciprocal space is essentially that expected from
Schr\"odinger dynamics~\cite{mark97a}, diffusing with mean standard
deviation of the packet size~\cite{sup}:
\begin{equation}
  \label{eq:viejun19120056MSK2015}
  \sigma_x(t)=\sqrt{\sigma_x^2(0)+({\hbar t}/[{2m_2\sigma_x(0)}])^2}\,.
\end{equation}
\begin{figure*}[!t]
  \includegraphics[width=1\linewidth]{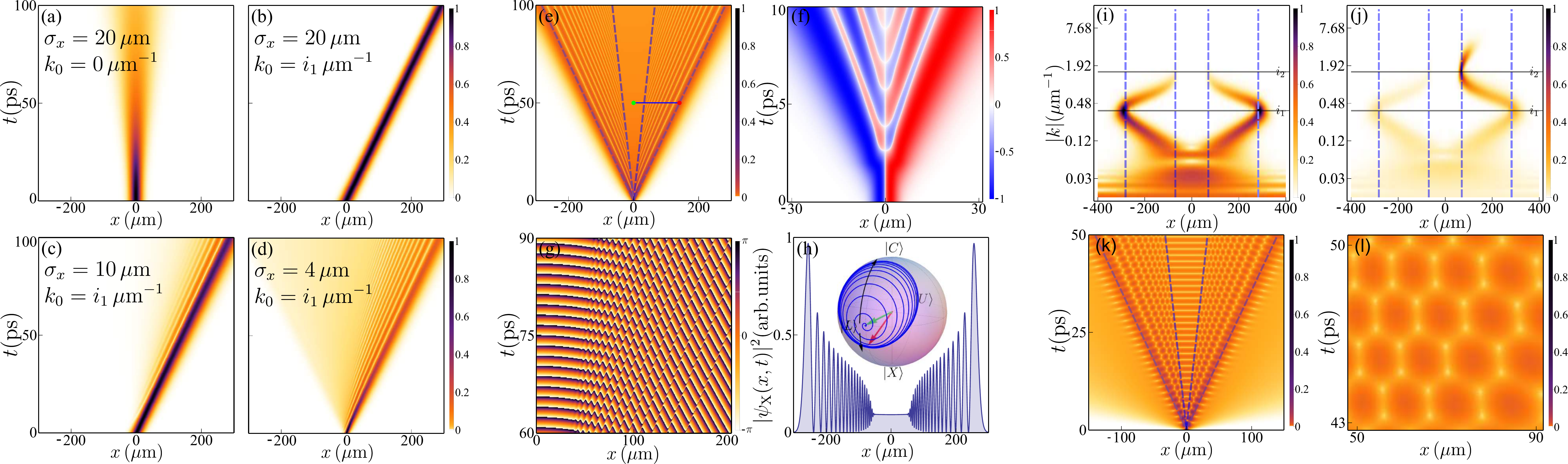}
  \caption{(Color online) (a--d) Propagation of lower polariton
    packets for various momenta and size, showing the emergence of the
    SIP for narrower packets. (e) Dynamics of an even narrower packet
    ($\sigma_x(0)=\unit{2}{\micro\meter}$ with no momentum
    $k_0=0$). (f) Current probability~$j$ at early times, showing the
    coexistence and interleaving of net counter-propagating flows. (g)
    phase map in a selected region, showing~$\pi$ jumps associated to
    each sub-peak. (h) intensity profile at
    $t=\unit{90}{\pico\second}$ with the evolution of the quantum
    state on the Bloch sphere corresponding to the path (from green to
    red) plotted in~(e). Wavelet decomposition of (e) at
    $t=\unit{100}{\pico\second}$, and (j) in the same configuration
    but exciting with a momentum $k_0=i_2$. (k) Spacetime honeycomb
    lattice when combining the SIP with Rabi oscillations by starting
    with a photon as an initial condition. (l) Zoom of the hexagonal
    lattice.}
  \label{fig:3}
\end{figure*}
Exciting one field only rather than eigenstate-superpositions (the
polaritons) yield Rabi oscillations.  Even in this simple case, there
are subtleties brought by these non-parabolic dispersions. In
particular, the degeneracy is lifted for some of the various concepts
of masses, famously unified for the gravitational and inertial masses
by Einstein as part of his theory of gravitation. For wavepackets,
there are two different effective masses $m_1$ and
$m_2$~\cite{larson05a}, describing respectively propagation and
diffusion. A wavepacket propagates with a group velocity
$v_\pm=\partial_k E_\pm(k)$. This defines the \emph{inertial mass}
$m_1$ that determines the wavepacket velocity from de Broglie's
relation $p=\hbar k$ and the classical momentum $p=m v_\pm$ as:
\begin{equation}
  \label{eq:m1}
  m_1(E,k)=\hbar^2 k(\partial_k E)^{-1}\,.
\end{equation}
A second mass $m_2$, that we will call it the \emph{diffusive mass},
is associated with the spreading of the wavepacket according to
Eq.~(\ref{eq:viejun19120056MSK2015}) and depends on the branch's
curvature; it reads:
\begin{equation}
  \label{eq:m2}
  m_2(E,k)=\hbar^2(\partial_k^2 E)^{-1}\,.
\end{equation}
These two masses $m_1$, $m_2$ and the packet velocity $v_-$ are
plotted in Fig.~\ref{fig:1}(b) for the Lower Polariton Branch
(LPB). Unlike parabolic dispersions, where they are equal, polariton
dispersions yield qualitatively differing inertial and diffusive
masses. In particular, the $k$-dependent inertial mass $m_1$ imposes a
maximum speed for the lower polaritons~\cite{sup}.  Beyond the
inflexion point~$i_1$, polaritons slow down if one increases their
momentum (at very large~$k$ the polariton becomes effectively a bare
particle again with no such kinematic restriction). The coupling of
the two fields with different masses results in the heavier one
lagging behind the other, as seen in Fig.~\ref{fig:2}(c) for the case
where a Gaussian photon wavepacket is imparted with a
momentum~$k=\unit{0.5/\hbar}{\micro\reciprocal\meter}$, achieved
experimentally by sending a pulse at an angle and overlapping both
branches.  This prevents the photon and exciton packets to propagate
Rabi-oscillating, and instead force a splitting in two beams---the
orthogonal polariton states which are eigenstates for the
corresponding wavevector, as shown by their trajectory on the Bloch
sphere---connected by a Rabi oscillating tunnel.  The Rabi
oscillations only take place when there is a spatial overlap between
the polaritons.  The two propagating packets maintain their coherence
despite their space separation and would Rabi oscillate if meeting
again, due, for instance, to a ping-pong reflection~\cite{anton13a}.
The splitting in two beams can be minimized by tuning parameters to
equalize the polaritons masses, in particular the inertial ones.
Combined with the bending of the Rabi oscillations in spacetime, which
can be achieved at nonzero detuning, this leads to propagation of Rabi
oscillations, that produce ultrafast subpackets moving inside a mother
packet, as shown in Fig.~\ref{fig:2}(d) and for three snapshots in
time in~(e). The subpackets, continuously formed in the tail of the
mother packet, propagate inside one order of magnitude faster, powered
by Rabi oscillations, before dying in the head.  Each sub-peak
acquires properties of an identifiable object that can be tracked in
time. The full dynamics is available in an accompanying
video~\cite{sup}.  Now on the diffusive mass~$m_2$: it diverges at the
two inflexion points~$i_{1,2}$ of the LPB and is negative in
between. Exciting at the inflexion points thus cancels diffusion of
the wavepacket as seen in Eq.~(\ref{eq:viejun19120056MSK2015}) and in
Fig.~\ref{fig:3}(a,b) with the propagation of a broad
($\sigma_x(0)=\unit{20}{\micro\meter}$) lower-polariton wavepacket
with an imparted momentum of (a)~$k_0=0$ and (b)~$k_0=i_1$.  The
excitation around the inflexion point has already been used to
generate bright solitons and soliton
trains~\cite{egorov09a,egorov10a,sich12a,sich14a}. In these cases, the
soliton mechanism is the conventional interplay between negative
effective mass and repulsive nonlinear interactions.  The role of the
high effective mass close to the inflexion point, which cancels the
diffusion, was not however fully estimated.

The interesting phenomenology discussed so far illustrates isolated
features of the polariton propagation.  A new physical picture emerges
when combining several aspects within the same wavefunction, leading
to the concept of self-interfering packets (SIP).  This is obtained
when reducing the size of the wavepacket in real space, that is,
increasing the staggering on the dispersion in momentum-space
($\sigma_k(0)=1/\sigma_x(0)$), to an extent enough to probe
polaritonic deviation from the parabolic dispersion. In this case, the
negative mass plays an explicit role.  Negative masses are a recurrent
theme in physics but this is typically meant for the inertial
mass~\cite{wimmer13a}. The sign of the diffusive mass would seem, at
first, not to play a role since it enters as a square in
Eq.~(\ref{eq:viejun19120056MSK2015}), and this is indeed the case for
a regular packet with momentum $i_1<k<i_2$. When straddling over the
divergence, however, self-interferences occur between harmonics of the
packet subject to the positive mass and others to the negative
mass. This result in a complete reshaping of the wavefunction, as
shown in Fig.~\ref{fig:3}(c,d) decreasing $\sigma_x(0)$ down to
$\unit{10}{\micro\meter}$ and $\unit{4}{\micro\meter}$. The part of
the packet that goes beyond the divergence is reflected back and
interferes with the rest of the packet that still propatages forward,
resulting in ripples. Reducing the packet to
$\sigma_x(0)=\unit{2}{\micro\meter}$ produces the striking pattern
seen in Fig.~\ref{fig:3}(e), without even the need of an imparted
momentum.  While for a parabolic dispersion, squeezing the packet in
space merely causes a faster diffusion, in the polariton case, there
is thus a critical diffusion beyond which the packet stops expanding
and folds back onto itself.  Since this happens when the wavefunction
encounters the inflection point of the polariton dispersion, there is
a ``mass wall'' against which the packet bounces back.  If the
dispersion also features another inflection point at larger~$k$, which
can be the case for small enough exciton masses, this reflection
happens again, this time resulting in a shielding from this
self-interference of the core of the mother packet, as shown on the
cut in intensity Fig.~\ref{fig:3}(h) (the diffusion cones are the
solution of $\partial_{k}^2E_-=0$, cf.~Fig.~\ref{fig:3}(e)).  More
importantly from a conceptual point of view, as a result of this
coexistence of masses of opposite signs within the same packet, the
mother wavepacket~$\ket{\psi}$ fragments itself into two trains of
daughter shape-preserving subpackets which travel in opposite
directions. The overall momentum $\bra{\psi}p\ket{\psi}=0$ is null but
the self-shaping of the wavefunction redistributes it through its
subpackets as a series of nonzero momenta.  Each sub-peak can be
identified by as a polariton as seen by following its quantum state on
the Bloch sphere, that lies onto the meridian between
$\ket{\mathrm{L}}$ and $\ket{\mathrm{X}}$, as shown in
Fig.~\ref{fig:3}(f) following a path (at $t=\unit{50}{\pico\second}$)
from the central aera---shielded from the self-interferences---to the
edge of the packet. The SIP can therefore be seen as a train of
successive polariton packets, ``emitted'' by the area shielded from
the self-interference at the rate of Rabi oscillations, and that
retain their individuality as they propagate inside the mother
packet. The full quantum state dynamics along these paths can also be
seen vividly in a supplementary video~\cite{sup}.  Successive peaks
furthermore feature a maximal phase shift of $\pi$ in the
phase~$\phi(x,t)$ of the total
wavefunction~$\psi(x,t)=|\psi(x,t)|\exp(i\phi(x,t))$, as shown in
Fig.~\ref{fig:3}(g).  Baring the fact that they do not involve
self-interactions to account for their cohesion and other properties
making them particles lookalike, these propagating subpackets behave
in many respects as soliton-like objects. The analogy with Airy beams
is conspicuous.

One can gain additional insights into the nature of the SIP through
the current probability
$j=i\hbar/2m_1(\psi^{\ast}\partial_x\psi-\psi\partial_x\psi^{\ast})$,
in Fig.~\ref{fig:3}(f) where the packet is plainly seen to alternate
backward and forward net flows. More sophisticatedly, considering the
wavelet transform (WT)~\cite{debnath_book15a}
$\mathbb{W}_{a,b}(\psi)=\frac{1}{\sqrt{|a|}}\int_{-\infty}^{+\infty}
\psi(x)\mathscr{G}^{\ast}\left(\frac{x-b}{a}\right)\, \mathrm dx$, in
our case, of the Gabor wavelet family
$\mathscr{G}(z)=\sqrt[\leftroot{-2}\uproot{2}4]{\pi}\exp(i \omega
x)\exp(-x^2/2)$, allows us to decompose the wavefunction into Gaussian
packets, which are the basic packets as far as propagation and
diffusion are concerned.  Such an extension of the Fourier transform
is common in signal processing but has found so far little echo to
study the dynamics of wavepackets~\cite{baker12a}.  We show in
Fig.~\ref{fig:3}(i) the energy density $|\mathbb{W}_{k,x}|^2$ of the
wavefunction in the $(x-k)$ plane at $t=100\mathrm{ps}$. One can see
clearly how the self-interferences force the polariton packet to
remain within the diffusion cone (blue dashed lines), by diverting the
flow backward, (i) one or (j) two times when the second inflexion
point ($k_0=i_2$) is reached.  Other fundamental connections could be
established. For instance, patterns strikingly similar to
Fig.~(\ref{fig:3}(e)) were observed in the quenched dynamics of a
quantum spin chains with magnons~\cite{liu14b}, a completely different
system. This suggests that coupled light-fields feature fundamental
and universal dynamical evolutions.  Combining this characteristic
pattern with that of Rabi oscillations leads to the space-time
propagation presented in Fig.~\ref{fig:3}(c). The protected area
simply exhibits Rabi oscillations. The outer area is propagating upper
polaritons and is not affected either by oscillations nor
interferences. In the SIP area, however, sitting between the two mass
walls, the interplay of Rabi oscillations and self-interferences
produces and hexagonal lattice. Here, instead of the emergence of
propagating particles, a spacetime crystal is formed with the manifest
ordering of the previously freely propagating train of
polaritons. This striking structure is, again, sculpted
self-consistently out of a simple Gaussian state by the dynamics of
coupled non-interacting fields.

In conclusion, we have shown the intricate wavepacket propagations of
coupled fields (polaritons). While the boundless diffusion of a
Schr\"odinger wavepacket in a parabolic dispersion ultimately leads to
complete indeterminacy, the polariton case can sustain traceable
objects with always well-defined properties, such as their shape,
position, momentum and quantum state. This gives rise to a concept of
particles similar to that brought by the soliton in nonlinear media or
Airy beams in non-interacting ones. While these are formed by
self-interaction and phase-shaping, the individuality of polaritons is
acquired and maintained through self-interferences powered by the Rabi
coupling. This shows that even in the linear regime, the polariton
dynamics is rich and able to produce intricate structures out of mere
Gaussian initial states.  This could lead to applications, by
imparting momentum powered by the Rabi oscillations or in the limit of
few particles, for quantum computing, by a proper wiring and directing
of the sub-packets, since all this happens in a strict linear regime.

\vfill\eject
\textbf{Acknowledgments:} This work is funded by the ``POLAFLOW'' ERC Starting
Grant. We thank L.~Dominici, D.~Ballarini, E.~del~Valle and
D.~Sanvitto for discussions.

%
%
%
\bibliographystyle{naturemag}
\bibliography{Sci,books,arXiv,david_bib}

\newpage

\onecolumngrid

\centerline{\large\textbf{Self-interfering wavepackets}}
\centerline{\large\textbf{Supplementary Material}}
\phantom{\ }

\twocolumngrid

\setcounter{equation}{0}
\renewcommand\theequation{S\arabic{equation}}

\section{Introduction and Outlook}

The theoretical model describing self-interacting wavepackets (SIP) is
simple: two coupled 1D Schr\"odinger equations in the linear regime
(see section~\ref{sec:miéjul1111733CEST2015} for the 2D case). The
simplicity of the equation is no guarantee that its extent and depth
are rapidly exhausted, as illustrated by the Schr\"odinger equation,
one of the most fundamental equation of modern physics, for which the
self-accelerating solution was discovered only in 1979. There should
be hope, however, that some closed-form solutions are available. In
Section~\ref{sec:miéjul1111811CEST2015} we show that the SIP, if it
has such analytical expressions, does not seem to be reducible to a
simple closed form. We could express it as a complex combinatorial
superposition of Bessel functions, as could be expected from
propagating packets, weighted by polaritonic factors, such as the
dressed momentum~$k_\Omega$. Before discussing this structure, we
start in Section~\ref{sec:miéjul1115153CEST2015} with more details on
the exact solutions, both from the formalism and numerical simulations
point of view, contrasting in particular the multitude of ways that
one has to poke at the polariton wavefunction. Last
Section,~\ref{sec:miéjul1115336CEST2015}, gives an overview of the
rest of the Supplementary material, that consists of animated movies
that illustrate, maybe better than equations, the mesmerizing dynamics
of polaritons.

\begin{figure}[th]
  \includegraphics[width=\linewidth]{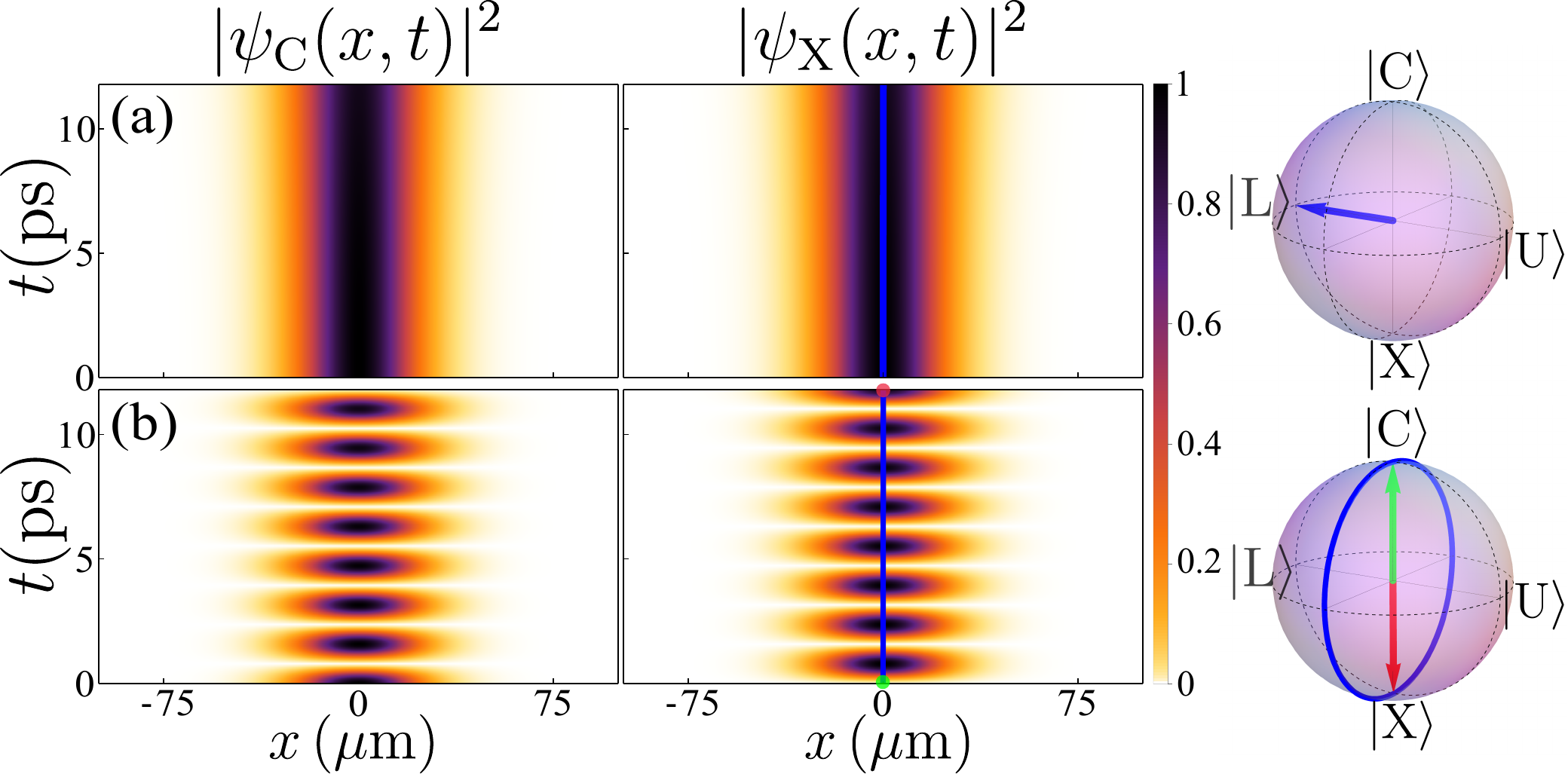}
  \caption{(Color online) Polariton propagation of delocalized
    wavepackets, as seen through the photonic~($\psi_\mathrm{C}$) and
    excitonic~($\psi_\mathrm{X}$) components for the cases of (a)
    upper row: a lower polariton at~$t=0$ and (b) lower row: a
    photon. The diffusion is negligible over the selected time window
    and the staggering on the dispersion too small to evidence
    polariton self-interference effects. Parameters: $\Omega_\mathrm{R}=2\, \textrm{meV}\, ,m_\mathrm{C}=0.025\, \textrm{meV}\,\textrm{ps}^2\,\mu\textrm{m}^{-2},
    m_\mathrm{X}=0.2\,\textrm{meV}\,\textrm{ps}^2\,\mu\textrm{m}^{-2}\,, \sigma_x=\unit{20}{\micro\meter}$.}
  \label{fig:S1}
\end{figure}

\section{More polariton propagation}
\label{sec:miéjul1115153CEST2015}

\begin{figure*}[t!]
  \includegraphics[width=.8\linewidth]{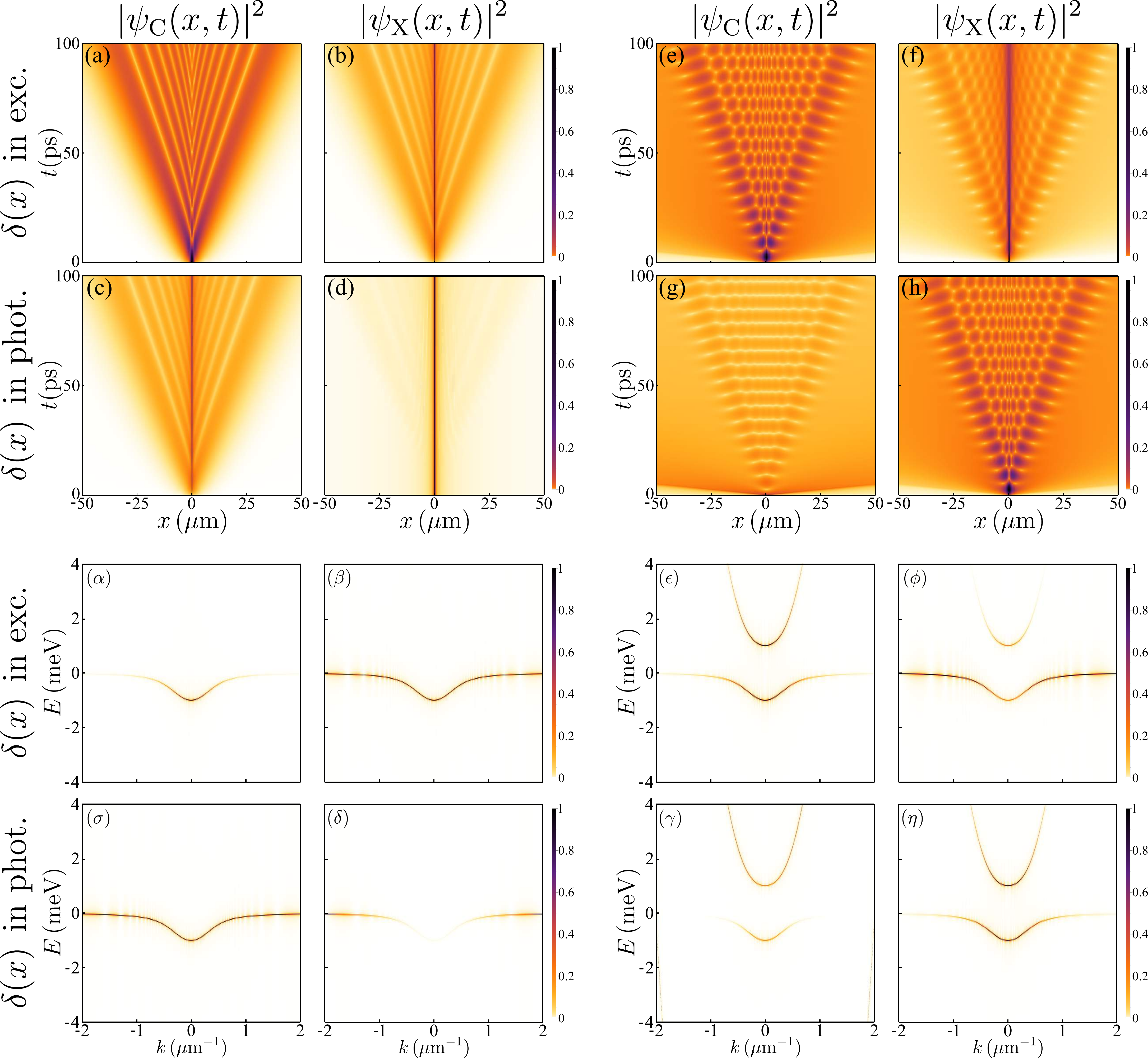}
  \caption{(Color online) Propagation of sharp packets. As seen in
    both real-space as a function of time (upper two rows) and through
    their dispersion (energy as a function of momentum, lower two
    rows) through the light emitted by the cavity~(C) or through the
    direct exciton emission~(X) for the cases of a lower polariton
    propagation (two columns on the left) and of a bare state
    propagation (two columns on the rigth). In the case of polaritons,
    (a--d), the exciton (upper row) or the photon (lower row) is
    perfectly localized at~$t=0$ with the other component defined such
    that the the particle remains on its branch. In the case of bare
    particles (e--h), the initial state is simply a photon (upper row)
    or an exciton (lower row) with the other component empty
    at~$t=0$. Greek-numbered panels correspond to the latin-numbered
    ones.  Parameters: $\Omega_\mathrm{R}=0.5\,\textrm{meV},
    m_\mathrm{C}=0.5\,\textrm{meV}\,\textrm{ps}^2\,\mu\textrm{m}^{-2}$.}
  \label{fig:S2}
\end{figure*}

The polariton propagator, Eq.~(1) of the main text is easily found
in~$k$ space (we work here with~$\Delta=0$):
\begin{widetext}
  \begin{equation}
    \label{eq:miéjun24181623CEST2015}
    \bra{k'}\Pi\ket{k}=
    \begin{pmatrix}
      \exp(\frac{-i k^2 m_+ t}{4})\left(\cos(\frac{k_\Omega^2 t}{4})+i \frac{k^2}{k_\Omega^2}m_- \sin(\frac{k_\Omega^2 t}{4})\right) & -\frac{i 4 \Omega_\mathrm{R}}{k_\Omega^2}\exp(\frac{-i k^2 m_+ t}{4})\sin(\frac{k_\omega^2 t}{4}) \\
      -\frac{i 4 \Omega_\mathrm{R}}{k_\Omega^2}\exp(\frac{-i k^2 m_+ t}{4})\sin(\frac{k_\omega^2 t}{4}) & \exp(\frac{-i k^2 m_+ t}{4})\left(\cos(\frac{k_\Omega^2 t}{4})-i \frac{k^2}{k_\Omega^2}m_- \sin(\frac{k_\Omega^2 t}{4})\right)\\
    \end{pmatrix}\delta(k-k')\,,
  \end{equation}
\end{widetext}
where we remind the important dressed momentum variable~$k_\Omega$:
\begin{equation}
  \label{eq:miéjun24100303CEST2015}
  k_\Omega\equiv\sqrt[4]{k^4 m_-^2+16m_+^2\Omega_\mathrm{R}^2}\,,
\end{equation}
with $m_{\pm}=(m_\mathrm{C} \pm m_\mathrm{X})/(m_\mathrm{C} m_\mathrm{X})$ (note again that $k_\Omega$ is a function of~$k$). We also assume $\hbar=1$. Polaritons are
maybe best formally defined as the states with a well-defined momentum
and, consequently, also energy. They satisfy:
\begin{equation}
  \label{eq:viejun26153857CEST2015}
  \Pi(t)\kket{k}_\pm=\exp(-iE_\pm t)\kket{k}_\pm\,,
\end{equation}
and as such are expressed as:
\begin{equation}
  \label{eq:miéjun24150811CEST2015}
  \kket{k}_\pm\propto
  \begin{pmatrix}
   E_\pm(k)\\
   1
  \end{pmatrix}\ket{k}\,,
\end{equation}
for~$+$ (resp.~$-$) the upper (resp.~lower) polariton, with~$E_\pm$
the pivotal polariton quantity, the dispersion:
\begin{equation}
  \label{eq:lunjul6163442CEST2015}
  E_\pm=k^2 m_+ \mp k_\Omega^2\,.
\end{equation}

\begin{figure}[t]
  \includegraphics[width=.9\linewidth]{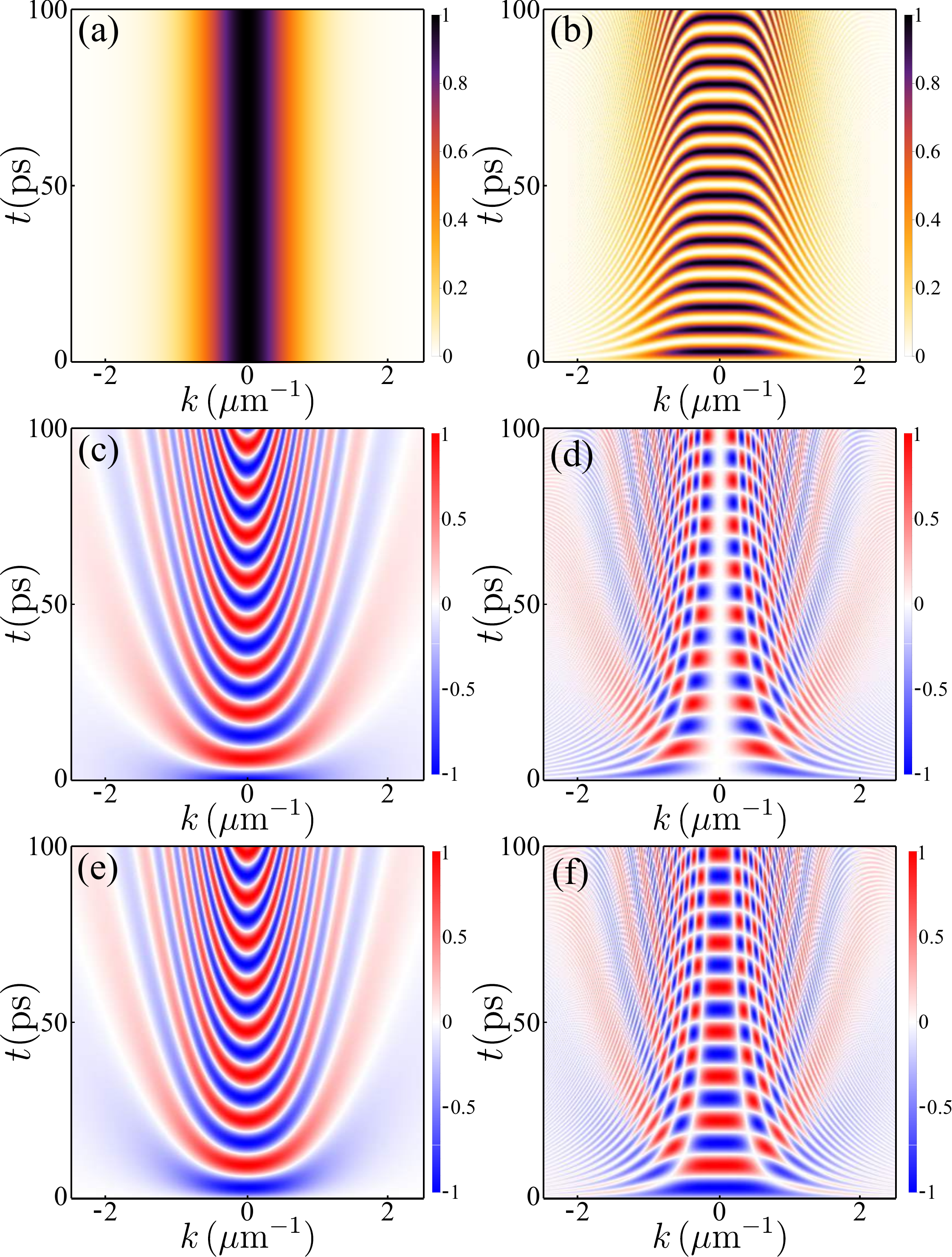}
  \caption{(Color online) Propagation in~$k$-space for non-diffusing
    packets of, left column, a lower polariton and, right column, a
    photon, as seen through, from up to bottom:
    $|\psi_\mathrm{C}(k,t)|^2$, $\Re(\psi_\mathrm{C}(k,t))$ and
    $\Im(\psi_\mathrm{C}(k,t))$. Parameters: $\Omega_\mathrm{R}=0.5\,\textrm{meV},
    m_\mathrm{C}=0.5\,\textrm{meV}\,\textrm{ps}^2\,\mu\textrm{m}^{-2}$.}
  \label{fig:S3}
\end{figure}

States~(\ref{eq:miéjun24150811CEST2015}) form a canonical
basis out of which a general polariton state is obtained by linear
combination:
\begin{equation}
  \label{eq:viejun26154721CEST2015}
  \kket{\psi}_\pm=\int_{-\infty}^{\infty}\phi_\pm(k)\kket{k}_\pm\,dk\,,
\end{equation}
with $\phi_\pm(k)$ the scalar-field (upper/lower) polariton
wavefunction. All the results in this text follow from the
impossibility to evolve such a general polariton state in time with
the complex rotation of free propagation as in
Eq.~(\ref{eq:viejun26153857CEST2015}), due to its two-component
character. We have already discussed in the main text some important
constrains of such a structure: for instance, that except for a
well-defined polariton state in~$k$-space, i.e., a completely
delocalized polariton in real space, the photon and exciton components
of a polariton cannot be jointly defined according to a given
wavepacket~$\phi(k)$, e.g., a Gaussian packet, since one component
gets modulated by the $E_\pm$ factor needed to maintain the particle
on its branch. Gaussian packets for both the photon and the exciton
result in populating both polariton branches. The general case
obviously admixes the two types of polaritons:
$\ket{\psi}=\sum_{\sigma=\pm}\int\phi_\sigma(k)\kket{k}_\pm\,dk$.
These results that impose strong constrains on a polariton wavepacket
must be contrasted with the conventional picture one has of the
polariton as a particle, which is that of states~$\kket{k}_\pm$ and
is, in good approximation, recovered for large enough packets as shown
in Fig.~\ref{fig:S1}. The particle is here broad enough in space to
have a small diffusion, cf.~Eq.~(2), and Rabi oscillations may be
present depending on the state preparation, which however do not
result in qualitative novelties. The situation is completely different
when considering overlaps in~$k$ space, that is, sharp packets in real
spaces. The propagation of such sharp packets is shown in
Fig.~\ref{fig:S2} for various initial conditions: as a lower polariton
(left part of the figure) or as a bare particle (exciton or photon,
right part of the figure) and such that the excitonic component is
perflectly localized at~$t=0$ (upper row) or the photon component is
(lower row). The wavepacket evolution is also shown as seen either
through its photon or exciton field. Experimentally, the photon field
is typically the one observed by recording the light leaked by the
cavity. One sees variations around the themes exposed in the main
text, with more or less pronounced features in some of the
configurations. The clearest effects are obtained for confined
excitons and observation through the cavity is always a good vantage
point. For each case of propagation in space-time, we also show in
Fig.~\ref{fig:S2}($\alpha$--$\eta$) the corresponding dispersion, which
is the double Fourier transform. This shows how, indeed, the lower
polariton only populates its own branch. It also provides an
alternative view of the various localized states, e.g., the localized
photon does not populate the photon-like part of the polariton branch
in the exciton spectrum (panel~$\delta$) and vice-versa with the
localized exciton in the photon spectrum (panel~$\alpha$).

We now proceed with the underlying Mathematical expressions. For, say,
the lower polariton case prepared so that the photon component is
perfectly localized at~$t=0$ (in the case where, for concision in the
notation, we assume from now on~$m_\mathrm{X}\rightarrow\infty$), we
find from
Eqs.~(\ref{eq:miéjun24181623CEST2015}--\ref{eq:viejun26154721CEST2015}):
\begin{subequations}
  \begin{eqnarray}
    \psi_\mathrm{C}(k,t)&=&\exp\left(-i\frac{k^2-k^2_\Omega}{4 m_\mathrm{C}}t\right)\frac{k^2_\Omega-k^2}{4 \Omega_\mathrm{R} m_\mathrm{C}}\,,\\
    \psi_\mathrm{X}(k,t)&=&\exp\left(-i\frac{k^2-k^2_\Omega}{4 m_\mathrm{C}}t\right)\,.\label{eq:lunjul6142711CEST2015}
  \end{eqnarray}
\end{subequations}
The result is extremely simple in this picture (time--momentum).
There is no time dynamics for the
density~$|\psi_{\mathrm{X},\mathrm{C}}|^2$: the perfectly localized
exciton at~$t=0$ results in a completely delocalized wavefunction at
all times~$|\psi_\mathrm{X}|^2=1$. The corresponding photon
wavefunction is qualitatively similar to a Voigt lineshape
in~$k$-space, which we will use in next Section to derive approximated
expressions. It is shown on the left column of Fig.~\ref{fig:S3}. In
all cases, however, there is of course a dynamics of the wavefunction
itself, as seen through its real and imaginary parts on the
figure. There is a slowing down of the oscillations with
increasing~$|k|$. The Fourier-transform in~$k$ of this pattern gives
the spacetime propagation in Fig.~\ref{fig:S2}(a).

The case of only the perfectly localized exciton as the initial
condition (i.e., in the photon vacuum rather than the field needed to
provide a lower polariton), i.e., $\psi_-(x,t=0)=(\delta(x),0)^T$,
is given directly by the columns of
Eq.~(\ref{eq:miéjun24181623CEST2015}):
\begin{widetext}
  \begin{subequations}
    \begin{eqnarray}
      \psi_\mathrm{C}(k,t)&=&\exp\left(-\frac{i k^2 t}{4 m_\mathrm{C}}\right)
      \left[\cos\left(\frac{k_\Omega^2t}{4m_\mathrm{C}}\right)-i\left(\frac{k}{k_\Omega}\right)^2\sin\left(\frac{k_\Omega^2t}{4m_\mathrm{C}}\right)\right]\,,
      \label{eq:2}\\
      \psi_\mathrm{X}(k,t)&=&\exp\left(-\frac{i k^2 t}{4 m_\mathrm{C}}\right)
      (-i\Omega_\mathrm{R}t)
      \operatorname{sinc}\left(\frac{k_\Omega^2t}{4 m_\mathrm{C}}\right)\,.
      \label{eq:3}
    \end{eqnarray}
  \end{subequations}
\end{widetext}
The corresponding propagation in momentum-space is shown on the right
column of Fig.~\ref{fig:S3}.  There is, this time, a dynamic in the
density, namely, the Rabi oscillations, with, in contrast to dynamics
of the real an imaginary parts, a speeding up of the oscillations with
increasing~$|k|$, corresponding to the effective Rabi frequency of
effectively detuned exciton-photon coupled states.  The
Fourier-transform in~$k$ of this pattern gives the spacetime
propagation in Fig.~\ref{fig:S2}(e).

\begin{figure}[t!]
  \includegraphics[width=.7\linewidth]{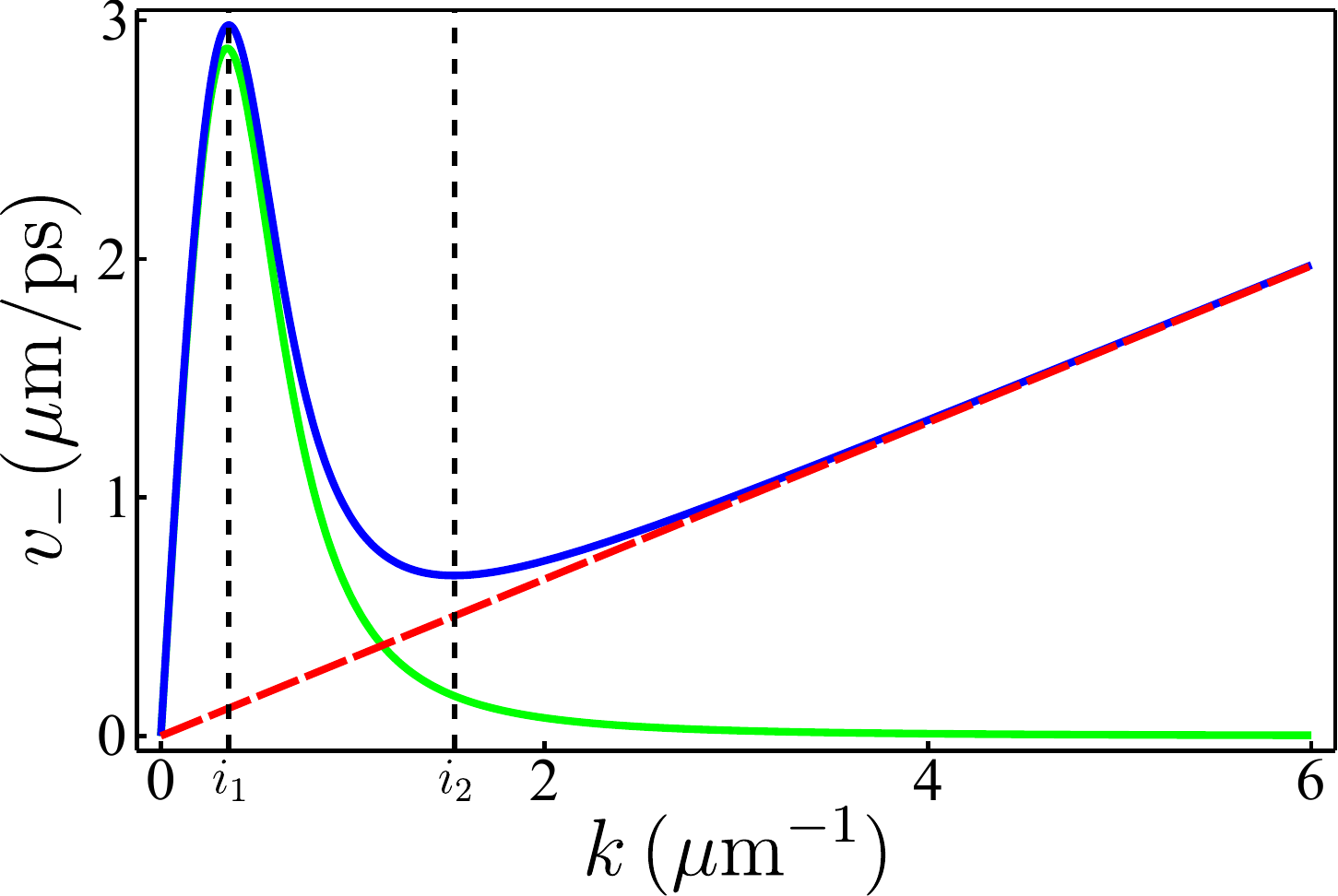}
  \caption{(Color online) Speed of the lower polariton as a function
    of its imparted momentum: there is a maximum speed for lower
    polaritons. Larger momenta result in slower particles. Local
    maxima are given by the inflexion points. At large enough~$k$, the
    polariton becomes exciton-like and suffers no such restriction. If
    the exciton mass is infinite, polaritons have an absolute maximum
    velocity.}
  \label{fig:S4}
\end{figure}

We have also discussed in the main text how a polariton wavepacket
propagates with a group velocity $v_\pm=\partial_k E_\pm(k)$ that, for
the lower polariton, features a local maximum. It reads:
\begin{equation}
\label{eq:lunjul6112841CEST2015}
v_-=\frac{k}{2 m_+}-\frac{\frac{\Delta}{k}-\frac{k}{2 m_-}}{ \sqrt{1-\frac{4 m_- \Delta}{k^2}+\frac{4 m_-^2 (4 \Omega_\mathrm{R}^2 +\Delta^2)}{k^4}}}\,,
\end{equation}
which is shown on Fig.~\ref{fig:S4} in blue.  The local maximum is
obtained at the first inflexion point of the dispersion, $k=i_1$,
where the polaritons also do not diffuse.  Increasing the momentum
makes the particle heavier and actually reduces its speed. A local
minimum is attained at the second inflexion point, $k=i_2$, where the
polariton also propagate without diffusion but now with a low speed.
At larger $k$, Eq.~(\ref{eq:lunjul6112841CEST2015}) becomes linear and
tends to the speed of the bare exciton, as indeed for $k\gg 0$,
$v_-(k)\to k/m_\mathrm{X}$, cf. Fig.~\ref{fig:S4} in dashed red for a
non detuned system.  If there is no second inflexion point (for an
infinitely heavy exciton mass), there is an absolute maximum speed for
the polaritons since the bare exciton group velocity vanishes (green
curve).

\section{Approximations}
\label{sec:miéjul1111811CEST2015}

\begin{figure}[t!]
  \includegraphics[width=\linewidth]{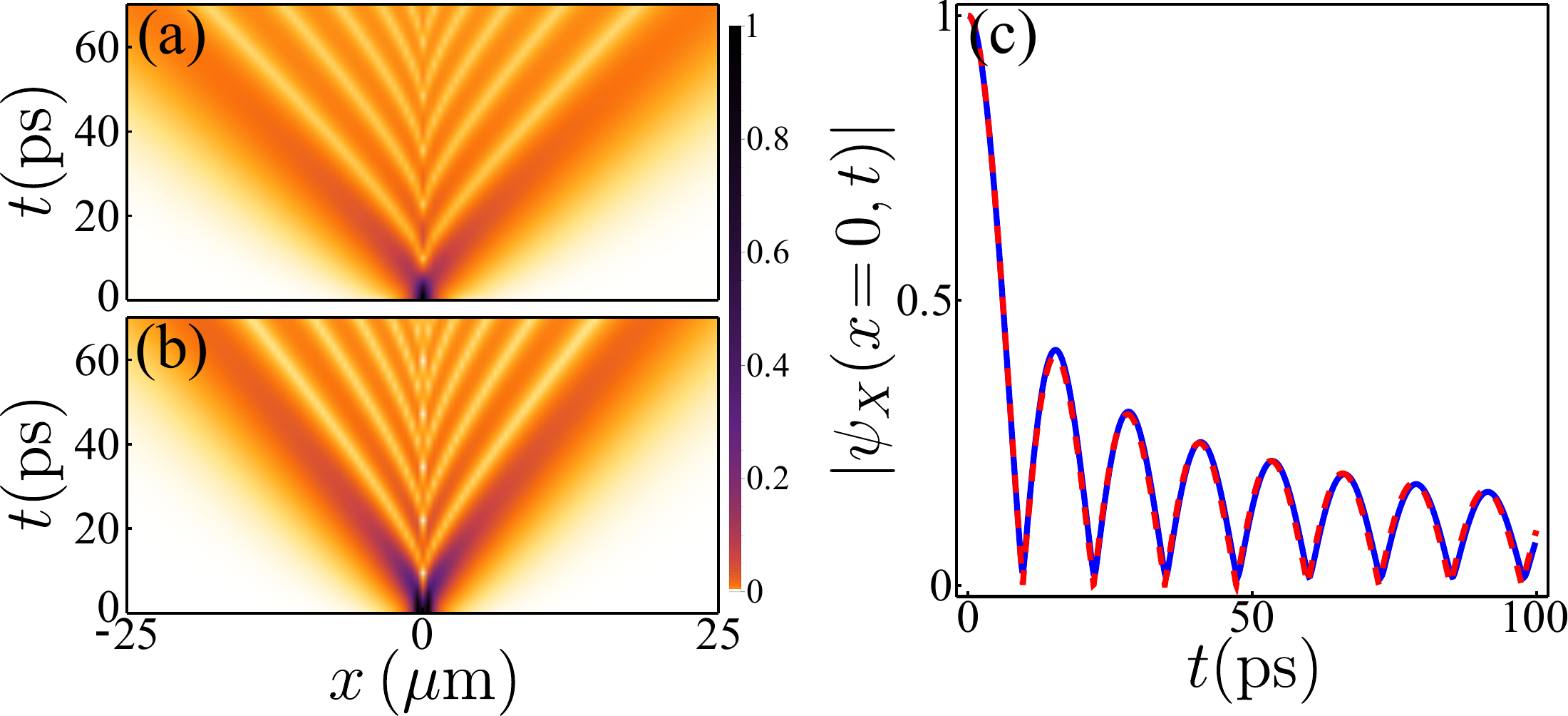}
  \caption{(Color online) (a,b) $|\psi_\mathrm{X} (x,t)|^2$ calculated
    from the exact result Eq.~(\ref{eq:lunjul6142711CEST2015}) and the
    approximated one, Eq.~(\ref{eq:7}). (c) Normalized intensity at
    the center of the wavepacket ($x=0$): exact result through the
    Fourier Transform of Eq.~(\ref{eq:lunjul6142711CEST2015}) (blue
    line) and the analytical expression Eq.\ref{eq:9} (red dashed
    line). The approximation appears to be exact at~$x=0$.}
  \label{fig:S5}
\end{figure}

We could not find manageable closed form expressions for the Fourier
Transform of the term $\exp(k_\Omega^2-k^2)$ that capture the (lower)
polaritonic self-interference effect.  The term $k^2-k_\Omega^2$ for
the (lower) polariton momentum distribution can however be well
approximated by a Voigt distribution, since it combines both
exponential and fat-tail types of decay.  Since the fat-tail is
expected to play a dominant role qualitatively, we assume simply a
Lorentzian distribution $f(k)={4\Omega_\mathrm{R}
  m_\mathrm{C}}/\big(1+\frac{k^2}{4 \Omega_\mathrm{R}
  m_\mathrm{C}}\big)$, thus approximating
Eq.~(\ref{eq:lunjul6142711CEST2015}) by:
\begin{equation}
\psi_\mathrm{X}(k,t)\simeq \frac{-\exp{\frac{i t \Omega_\mathrm{R}}{1+k^2/(4 m_\mathrm{C} \Omega_\mathrm{R})}}}{1
+k^2/(4 m_\mathrm{C} \Omega_\mathrm{R})}\,,
\label{eq:6}
\end{equation}
\\
which Fourier Transform can be obtained by a series expansion of the
exponential, providing the real-space dynamics of the SIP:
\\
\begin{multline}
  \psi_\mathrm{X}(x,t)\simeq
  \\ \sum_{n=0}^{\infty}\frac{-4\sqrt{\pi} (m_\mathrm{C} \Omega_\mathrm{R})^{\frac{3+2n}{4}} (i t \Omega_\mathrm{R})^n |x|^{\frac{1}{2}+n} K_\alpha (2\sqrt{\Omega_\mathrm{R} m_\mathrm{C}} |x|)}{n!^2}\, ,
\label{eq:7}
\end{multline}
where $K_\alpha(z)$ is the modified Bessel functions (solution of th
equation $z^2y''+z y' -(z^2 +n^2)y=0$) and with $\alpha=n+1/2$.  The
propagation of the wavepacket calculated numerically with the Fourier
Transform (a) and the one obtain with the analytical formula in
Eq.~\ref{eq:7} (b) are compared in Fig.~\ref{fig:S5}.  There is a good
qualitative agreement, with only the size of the propagation cone
differing slightly, due to the width difference between the Lorentzian
distribution and the actual one. One can therefore trust the
approximation to give some insights into the nature of the SIP.
First, the SIP is indeed a phenomenon of many interferences.  The
convergence of the series is obtained for a number of terms in the sum
that increases linearly with time~$t$, showing how each new peak
arises from an added term and thus a next order in the
interference. Also, at the center of the wavepacket ($x=0$), the
previous expression can be reduced to a simple form:
\begin{equation}
  \psi_\mathrm{X}(x=0,t)\simeq -2\pi\sqrt{\Omega_\mathrm{R} m_\mathrm{C}}\e^{\frac{it \Omega_\mathrm{R}}{2}} J_0\left(\frac{t \Omega_\mathrm{R}}{2}\right)\,,\\
\label{eq:9}
\end{equation}
where $J_n(z)$ is the Bessel function of the first kind. Since the
departure between Eq.~(\ref{eq:7}) and the exact numerical solution is
mainly due to the extent of the envelope of the momentum, one can
expect a better agreement at~$x=0$, and indeed we find that there is a
perfect match, as seen in Fig.~\ref{fig:S5}(c). Equation~(\ref{eq:9})
confirms that the successive peaks that shape the SIP appear at the
Rabi frequency~$\Omega_\mathrm{R}$. The photon mass $m_\mathrm{C}$ (we
remind that we assumed here an infinite exciton mass) acts only on the
intensity.  In the same way, one can obtain the corresponding series
for the photon field:
\begin{multline}
\psi_\mathrm{C}(x,t)\simeq
\\ \sum_{n=0}^{\infty}\frac{4\sqrt{\pi} (m_\mathrm{C} \Omega_\mathrm{R})^{\frac{2n+1}{4}} (i t \Omega_\mathrm{R})^n |x|^{n-\frac{1}{2}} K_\beta (2\sqrt{\Omega_\mathrm{R} m_\mathrm{C}} |x|)}{n! \Gamma(n)}\,,
\label{eq:10}
\end{multline}
with $\beta=n-1/2$, showing how both structures are tightly related
and the sort of complexity that describes them.  Series for photons or
excitons as initial conditions can also be obtained from
Eq.~\ref{eq:2} and \ref{eq:3}, involving Hypergeometric functions
which, however, are too cumbersome to be written here.

\section{SIP in two dimensions}
\label{sec:miéjul1111733CEST2015}
\begin{figure}[t!]
  \includegraphics[width=.75\linewidth]{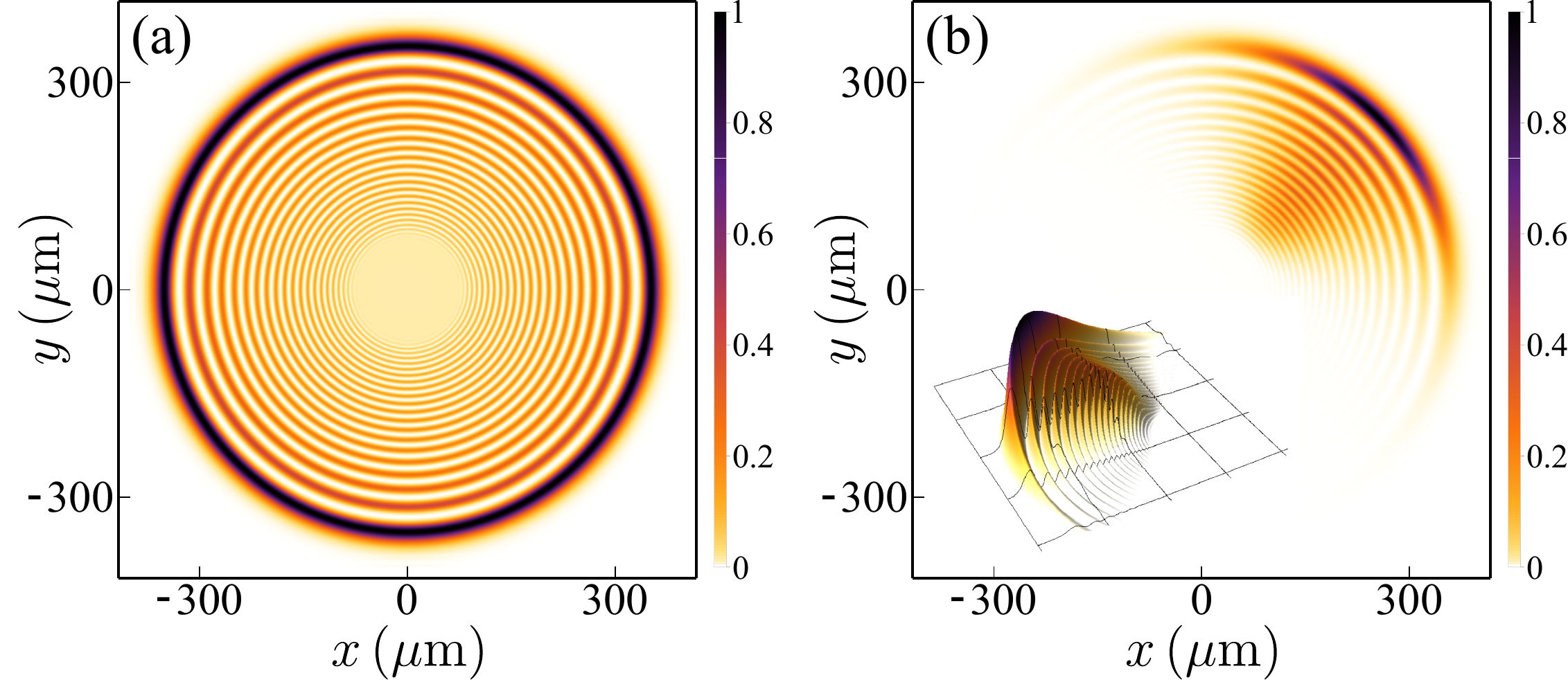}
  \caption{(Color online) SIP in two-dimensions. (a) With full radial
    symmetry when self-interfering in all directions. A central disk
    is shielded from interference just like in the 1D case and feature
    a flat plateau of lower polaritons. (b) When imparted with
    momentum, the packet self-interfere in the direction of its
    propagation and diffuses radially. Parameters : $\Omega_\mathrm{R}=2\,\textrm{meV}\, ,m_\mathrm{C}=0.025\, \textrm{meV}\,\textrm{ps}^2\,\mu\textrm{m}^{-2}, m_\mathrm{X}=0.2\,\textrm{meV}\,\textrm{ps}^2\,\mu\textrm{m}^{-2}, $ (a)  $\sigma_x=\unit{2}{\micro\meter}$, (b) $\sigma_x=\unit{4}{\micro\meter}, \, k_x=k_y=i_1$.}
  \label{fig:S6}
\end{figure}
So far, we have considered 1D cases, which are indeed possible in
heterostructures by confining in the two other dimensions (quantum
wires). Polariton propagation is however popular in the 2D geometry as
well and we quickly discuss what happens in this configuration.  Since
the system is linear and uncoupled in transverse coordinates, the
dynamics follows trivially from the previous results and symmetry.
With a full radial symmetry, all the phenomenology is conserved in~2D,
with rotational invariance.  This is shown in Fig.~\ref{fig:S6}(a)
where the propagation of a SIP is shown after
$\unit{100}{\pico\second}$, after preparing a narrow Gaussian
wavepacket at~$t=0$ (and at the origin of the plane). This is the
counterpart of the case of Fig.~3(e) of the main text. The propagation
and thus the shape of the wavepacket are similarly determined by the
dispersion and the way it is excited. There is also, depending on the
proximity of the second inflection point, a central area shielded from
the interferences, which is now a disk, whose diameter is also
determined by a zero of~$\partial_k^2E_-=0$. Also, like in
the 1D case, one can impulse the propagation of the packet in a
desired direction by imparting a momentum. This is shown in
Fig.~\ref{fig:S6}(b) with the propagation of a smaller wavepacket when
exciting the dispersion at the first inflexion point ($k_x=k_y=i_1$).
By using squeezed Gaussian packets, one can propagate a SIP with
fronts that remain parallel and orthogonal to the direction of motion.

\section{Movies}
\label{sec:miéjul1115336CEST2015}

We also provide two movies that illustrate vividly the polariton
dynamics.  

The movie \texttt{I-QuantumState.avi} animates Fig.3~(e,h) of the main
text. It shows how a narrowly squeezed lower polariton wavepacket
self-interferes and produces, as a result, two trains of sub-packets
propagating back to back, emerging from a polariton sea shielded from
the interferences. To show how the overall structure of the SIP is
connected to the individual identity of each packet, we also present
dynamically on the Bloch sphere the evolution of the quantum state on
a path that links the center of the wavepacket to the side of the
diffusion cone (from the green to the red point on the density plot,
mached with the green and red arrows on the sphere). The state in the
central area---shielded from interferences---is the lower
polariton. In the interferences zone, crossing a fringe induces a loop
on the sphere that crosses the meridian of states
$\ket{\mathrm{C}}$--$\ket{\mathrm{L}}$--$\ket{\mathrm{X}}$.  Each peak
converges in time towards a well defined polariton state on the
meridian.

The movie \texttt{II-PolaritonRiffle.avi} animates the case of
Fig.~2(d,e) of the main text. It shows the propagation of a photon
wavepacket (at $t=0$) with a momentum and negative detuning. A
judicious choice of parameters permits to maintain a spatial overlap
between the two polaritons, conserving the Rabi oscillations which,
due to the detuning, are bent in spacetime (see main text).  This
results in the propagation of ultrafast subpackets within the mother
packet. This is well seen on the animation after
\unit{35}{\pico\second} of animation time, which is the time needed
for the packet to develop the structure (the initial condition is a
Gaussian packet).

\end{document}